\documentclass[aps,twocolumn,prd,nofootinbib,showpacs]{revtex4}
\usepackage[pdftex,bookmarks]{hyperref}

\usepackage{amsfonts}
\usepackage{amsmath,amssymb}
\usepackage[T1]{fontenc}
\usepackage{mathrsfs}
\usepackage{epsfig,epstopdf}

\usepackage[absolute]{textpos}

\usepackage{graphics,wrapfig,times}
\usepackage{graphicx}

\usepackage{color}
\definecolor{D}{rgb}{0.00,0.17,0.48}
\definecolor{M}{rgb}{0.00,0.02,0.83}
\definecolor{L}{rgb}{0.58,0.79,1.00}
\definecolor{R}{rgb}{1,0,0}
\definecolor{B}{rgb}{0.00,0.00,0.00}
\definecolor{P}{rgb}{0.00,0.30,0.60}
\definecolor{W}{rgb}{1,1,1}

\TPGrid[40mm,40mm]{15}{25}  

 \makeatletter
 \newcommand{\pback}[1]{{
   \let\@rrow=\leftarrowfill
   \mathchoice{\AIN@stemPullBack{#1}{\@rrow}}{\AIN@stemPullBack{#1}{\@rrow}}
     {\AIN@indxPullBack{#1}{\@rrow}}{\AIN@indxPullBack{#1}{\@rrow}}}
   \vphantom{#1}}

 \newcommand{\AIN@stemPullBack}[2]{
   \vtop{\mathsurround=0pt
   \ialign{##\crcr$\textstyle{#1}\strut$\crcr
     \noalign{\kern-0.4ex\nointerlineskip}{\tiny#2}\crcr}}}

 \newcommand{\AIN@indxPullBack}[2]{
   \vtop{\mathsurround=0pt
   \ialign{##\crcr\hfil$\scriptstyle{#1}$\hfil\crcr
     \noalign{\kern+0.4ex\nointerlineskip}{\tiny#2}\crcr}}}

\def\bar{\overline}
\def\be{\begin{equation}}
\def\ee{\end{equation}}
\def\bea{\begin{eqnarray}}
\def\eea{\end{eqnarray}}
\def\ba{\begin{array}}
\def\ea{\end{array}}

\def\={\hateq}
\def\puto#1{\rlap{\raise.5ex\hbox{\char'27}}{#1}}

\newcommand{\nn}{\nonumber}
\newcommand{\half}{\frac{1}{2}}
\def\eth{\text{\dh}}
\def\thorn{\text{\th}}
\def\a{\alpha}
\def\b{\beta}
\def\c{\gamma}
\def\d{\delta}

 \def\tt2{{\tilde{2}}}

\def\.{\cdot}
\def\D{{\cal D}}

\def\L{{\cal L}}

\def\Re{{\rm Re}}
\def\Im{{\rm Im}}
\def\l{\ell}

\def\be{\begin{equation}}
\def\ee{\end{equation}}
\def\bea{\begin{eqnarray}}
\def\eea{\end{eqnarray}}
\def\ba{\begin{array}}
\def\ea{\end{array}}

\def\up{\stackrel}

\newcommand{\eqhat}{\mathrel{\widehat\mathalpha{=}}}
\def\={\eqhat}

\begin{document}

\title{Gravitational radiation and angular momentum flux from a slow rotating dynamical black hole}

\author{Yu-Huei Wu}\email{yhwu@phy.ncu.edu.tw}
\affiliation{1. Center for Mathematics and Theoretical Physics, National Central University, Chungli, 320, Taiwan.\\
2. Department of Physics, National Central University, Chungli, 320, Taiwan.}

\author{Chih-Hung Wang}\email{chwang1101@phys.sinica.edu.tw}
\affiliation{1. Department of Physics, Tamkang University, Tamsui, Taipei 25137, Taiwan.\\
 2. Institute of Physics, Academia Sinica, Taipei 115, Taiwan.\\
3. Department of Physics, National Central University, Chungli, 320, Taiwan.}

\begin{abstract}

A four-dimensional asymptotic expansion scheme is used to study the next order effects of the nonlinearity near a spinning dynamical black hole.
The angular momentum flux and energy flux formula are then obtained by constructing the reference frame in terms of the compatible constant spinors and the compatibility of the coupling leading order Newman-Penrose equations.
By using the slow rotation and small-tide approximation for a spinning black hole, we chose the horizon cross-section is spherical symmetric. It turns out the flux formula is rather simple and can be compared with the known results.
Directly from the energy flux formula of the slow rotating dynamical horizon, we find that the physically reasonable condition on the positivity of the gravitational energy flux yields that the shear will monotonically decrease with time. Thus a slow rotating dynamical horizon will asymptotically approaches an isolated horizon during late time.

\end{abstract}

 \maketitle

\section{Introduction}


Null infinity and black hole horizon have similar geometrical properties. They are both three dimensional hypersurfaces and have the gravitational flux across them. The physical properties of null infinity can be studied in the conformal spacetime with finite boundary. Thus the conformal method provides an alternative way to study Bondi-Sachs gravitational radiation near null infinity, which was first proposed by Penrose \cite{Penrose-86}.
The boundary of a black hole is asymptotically non-flat and one may not be able to apply the conformal method to study the boundary problem of a dynamical black hole.  Rather than using the symmetry for the whole space-time to locate the boundary of a black hole, Ashtekar \textit{et al.} use a rather mild condition on the symmetry of the three dimensional horizon \cite{Ashtekar99b, Ashtekar02}. This quasi-local definition for the black hole boundary makes it possible to study the gravitational radiation and the time evolution of the black hole.

In this paper, we use the Bondi-type coordinates to write the null tetrad for a spinning dynamical horizon (DH). The boundary conditions for the quasi-local horizons can be expressed in terms of Newman-Penrose (NP) coefficients from the Ashtekar's definition on DH.
Unlike Ashtekar \textit{et al}'s \cite{Ashtekar99b, Ashtekar02} three dimensional analysis, we adopts a 4-dimensional asymptotic expansion to study the neighborhoods of generic isolated horizons (IHs) and dynamical horizons (DHs). Since the asymptotic expansion has been used to study gravitational radiations near the null infinity \cite{NU, ENP}, it offers a useful scheme to analyze gravitational radiations approaching another boundary of space-time, black hole horizons. We first set up a null frame with the proper gauge choices near quasi-local horizons and then expand Newman-Penrose (NP) coefficients, Weyl, and Ricci curvature with respect to radius. Their fall-off can be determined from NP equations, Bianchi equations, and exact solutions, e.g., the Vaidya solution.  This approach allows one to see the next order contributions from the nonlinearity of the full theory for the quasi-local horizons.

We have shown that the quasi-local energy-momentum flux formula for a non-rotating DH by using asymptotic expansion yields the same result as Ashtekar-Krishman flux \cite{Wu2007, WuWang-PRD-2009}. For slow rotating DH, we have presented our results in \cite{Wu2007}, however, we use an assumption of vanishing NP coefficient $\lambda$ on DH.  Furthermore, the energy-momentum flux formula has a shear (NP coefficient $\sigma$) and a angular momentum (NP coefficient $\pi$) coupling term. Since it is unclear whether the existence of this term carries any physical meaning or it may due to our assumptions, 
we thereby extend our previous work on IHs and DHs into a more general case. 



An algebraically general structure (Petrov type I) of spacetime is thought to be related with gravitational radiation for an isolated source and can tell us more about the inner structure of the gravitating source. The Weyl scalars $\Psi_k, k=0,..,4$ can be expanded in terms of an affine parameter $r$  along each outgoing null geodesic based on assumption of compatification of null infinity \cite{Sachs61, Penrose-86}. Here $\Psi_k=O(r^{k-5}), k=0,..,4$ and one may find that it peels off more and more when moving inward along null ray. From Ashtekar's definition of isolated horizon, it implies that $\Psi_0,\Psi_1$ vanishes on horizon. Therefore the space-time is algebraically special on isolated horizon. However, spacetime may not be algebraically special for an arbitrary DH.
The corresponding peeling theorem for an arbitrary DH is crucial for our gravitational radiation study.  Due to the difficulties of knowing the fall-off of Weyl scalars, we use Kerr-Vaidya solution to serve as our basis for choosing the fall-off of Weyl scalars,which is $\Psi_0,\Psi_1$ vanishing on DH, in our previous work on slow-rotating DH \cite{Wu2007}. So spacetime structure on slow-rotating DH is still assumed to be algebraically special. However, according to the gravitational plane-wave solutions, $\Psi_0$ and $\Psi_4$ indicate the ingoing and outgoing gravitational waves, respectively. It seems physically unsatisfactory to assume $\Psi_0$ vanishing on DH. Moreover, the algebraically general space-time allows four roots of the equation, which correspond to the principal null-directions of Weyl scalars, and describes the gravitational radiation near the gravitating source. Therefore it would be more reasonable for one to consider an algebraically general space-time on an evolving DH. From the reduction and the decoupling of the equations governing the Weyl scalars, instead of assuming $\Psi_0,\Psi_1=0$ on DH, we set $\Psi_1, \Psi_3$ vanishing on a spinning DH. This is a similar setting with the perturbation method (Also see Chandrasekhar \cite{Chandrasekhar}).

We present the results of asymptotic expansion for a spinning DH in Sec. \ref{AEDH}. However, it maybe too general to yield
some interesting physical results. By considering the small-tide and slow rotate of DH and using slow rotate Kerr solution as a basis, we use two sphere conditions of DH cross section for our later calculation.
The NP coefficient $\lambda_0$ (shear for the incoming null tetrad $n$) on DH is no longer assumed to be vanished when calculating the flux formula. The index $0$ on NP coefficients denotes their values on DH. Directly from non-radial NP equations, we find that $\sigma_0$ and $\pi_0$ coupling terms can be transformed into $\pi_0$ terms only, so the problems of our previous work \cite{Wu2007} are resolved.


Though the exact solution for a stationary rotating black hole has been found near fifty years, the spacetime with rotation remains its ambiguity and difficulty for quasi-local mass expressions and boundary condition. For example, the existence of angular momentum will not change the boundary condition for the null infinity, however, it will affect the boundary condition of  a black hole.
Among the well-known quasi-local mass expressions named Komar, Brown-York and Dougan-Mason, only Komar integral of the quasi-local mass for an arbitrary closed two surface can go back to the unique Newtonian quasi-local mass \cite{WuWang-2008-a}. Unfortunately, Brown-York and Dougan-Mason mass can return to the unique surface integration of the Newtonian mass in the covariant Newtonian spacetime only for the spherically symmetry sources.
In GR, quasi-local mass expressions for Kerr solution disagree one another \cite{Bergqvist92}. Different quasi-local expressions give different values of quasi-local mass for Kerr black hole. At null infinity, there is no generally accepted definition for angular momentum \cite{Szabados-04}. Unfortunately, no explicit expression for Bramson's angular momentum in terms of the Kerr parameters $m$ and $a$ is given \cite{Bramson75a}. We use Komar integral to calculate angular momentum since it gives exactly $m a$ for Kerr solution. Although different quasi-local expressions yield the different results for Kerr solution, our main motivation is to analyze and discuss the compatibility of the coupling NP equations from asymptotic expansions. We both calculate quasi-local mass and flux for a spinning DH based on two spinors (Dougan-Mason) and Komar integral. It is found that these two expressions yield the same result.



Bondi and Sachs use no-incoming radiation condition for gravitational wave on null infinity \cite{BBM62, Sachs61}. However, no-incoming radiation condition is only true for linearized theories, e.g., electrodynamics and linearized GR, as to exclude the incoming rays. The incoming pulse waves do not destroy the asymptotic conditions for null infinity since they are admitted by formalism. Their existence may play an important role in the interpretation of the new conserved quantities (NP constants) \cite{NP65, NP68}. The interpretation and physical meaning of these constants have been a source of debate and controversy until today. Some physical discussions and application of them can be found in \cite{Dain-Kroon-02, Jaramillo-Kroon-Gourgoulhon-08}. Despite the vagueness of the physical meaning of these conserved quantities, in the full nonlinear gravitational theory, the mass and momentum are no longer absolutely conserved and can be carried away by the outgoing gravitational wave, so as to give a positive energy flux at infinity.  Here we consider a spacetime inner boundary, e.g, a spinning DH in this paper. By the aid of using asymptotic constant spinor to define spin frame as the reference frame for our observation, mass and angular momentum flux can be calculated. According to the coupling NP equations from the asymptotic expansion analysis, such a system will gain energy and will cause the radius of the black hole to increase. From similar argument, the outgoing waves \textit{do not} change the boundary conditions of the quasi-local horizons (DHs or IHs) and make no contribution to flux, while incoming wave will cross into DHs. The existence of incoming wave indicates the difference between IHs and DHs. The mass and momentum are carried into black hole by the incoming gravitational wave.

In the Ashtekar-Kirshnan's 3-dimensional analysis, it gives no constraints on $\sigma_0$'s time evolution. However, through our 4-dimensional asymptotical expansion scheme, we observe that the $\sigma_0$ will monotonic decrease with time once the positivity of gravitational energy flux is hold on a slow rotating DH. It means that the slow rotating DHs will gradually settle down to IHs as $\sigma_0$ approach zero, which is physically reasonable. It is similar to a physical assumption saying that the mass loss cannot be infinite large for null infinity \cite{NU}. However, rather than assuming that mass loss cannot be infinite large we obtain this result directly from asymptotic expansion analysis for a slow rotating DH together with the physical arguments.
Further from the commutation relations, we find that the horizon radius of a slow rotating DH will not accelerate. The slow rotating dynamical horizon increases with a constant speed.
There is one more interesting point about the peeling properties for a slow rotating DH. It is known that the peeling properties refer to different physical asymptotic boundary conditions of a slow rotating black hole. By comparing our current work to previous one \cite{WuWang-PRD-2009}, which have different peeling properties, and also due to the monotonic decrease of $\sigma_0$, we propose that the setting of Weyl scalars in this work makes it excludes the possibility to absorb the gravitational radiation from near by gravitating source.

The plan of this paper is as follows. In Sec. \ref{DH}, we review the definition of DH and express Ashtekar-Krishnan's 3-dimensional analysis of DH in terms of NP coefficients.
The gauge choices and boundary conditions of a spinning DH are applied to the asymptotic expansion in Sec. \ref{AEDH}. In Sec. \ref{Slow}, we first examine the gauge conditions of slow rotating Kerr solution in Subsec. \ref{S-Kerr}. Later we use the two sphere condition for a slow rotating DH with small tide in Subsec. \ref{sphere}. The results of asymptotic expansion are largely simplified by considering DH's cross section as a two sphere. Angular momentum and its flux for a slow rotating DH are calculated by using Komar integral in Sec. \ref{J-DH}. Energy-momentum and its flux of a slow rotating DH are obtained in Sec. \ref{EM-spinningDH}. We first calculate mass and mass flux by using Komar integral in Subsec. \ref{K-mass}. Then, mass and mass flux of a slow rotating DH is calculated by using two spinor method in Subsec. \ref{Mass-spinor}. The time evolution of shear flux  and its monotonic decrease is discussed here. We find that either Komar integral or two spinor method yields the same result.

In this paper, we adopt the same notation as in  \cite{Ashtekar99b, Ashtekar02} for describing generic IHs and DHs. However, we choose the different convention $(+---)$, which is a standard convention for the NP formalism \cite{Penrose-84}.  The necessary equations, i.e., commutation relations, NP equations and Bianchi identities, for asymptotical expansion analysis can be found in p. 45-p. 51 of \cite{Chandrasekhar}. We use "$\hat{=}$" to represent quantities on a dynamical horizon (ignore $O(r')$) and use "$\cong$" to represent quantities on a slow rotating horizon (ignore $O(a^2)$).


\section{Ashtekar dynamical horizon} \label{DH}

\subsection{The dynamical horizon }

The generic IHs are taken as the equilibrium state of the
DHs. The DHs can be foliated by
marginally trapped surface S. Therefore, the expansion of the
outgoing tetrad vanishes on DHs.

\medskip

\paragraph{Definition} A smooth, 3 dimensional, space-like sub-manifold
$H$ of space-time is said to be a \textit{dynamical horizon} (DH)
if it can be foliated by a family of closed 2-manifold such that:
(1) on each leaf, $S$, the expansion $\Theta_{(\l)}$ of one null
normal $\l^a$ vanishes, (2) the expansion $\Theta_{(n)}$ of the
other null normal $n^a$ is negative.

\medskip

From this definition, it basically tells us that a dynamical horizon is
a space-like hypersurface, which is foliated by closed, marginally
trapped two surface. The requirement of the expansion of the
incoming null normal is strictly negative since we want to
study a black hole (future horizon) rather than a white hole.
Also, it implies
\bea \Re [\rho] \hat{=}0, \Re[\mu]<0. \label{DHgc}\eea

\subsection{Dynamical horizon in terms of
Newman-Penrose coefficients}

If we contract the stress energy tensor with a
time-like vector, then in components $T^k\;_0$ represents the
energy flux of matter field. Therefore we can use a time-like
vector $T^a$ and contract it with the stress energy tensor to
define the flux of the matter energy. Here we are more interested
in the energy of the matter field associated with a null
direction. One can thus calculate the flux of energy associated
with $\xi^a = N \l^a$. The flux of matter energy across $H$ along
the direction of $\l$ is given by
\bea F_{\textrm{matter}} := \int_H T_{ab} T^a \xi^b d^3 V.
\label{matterflux}\eea
The dynamical horizon is a space-like surface, the Cauchy data $(
\;^{(3)}q_{ab}, K_{ab})$ on the dynamical horizon must satisfy the
scalar and vector constraints
\bea H_S:  &=& ^{(3)} R + K^2 -K_{ab}K^{ab} = 16\pi T_{ab}T^a T^b, \label{Scalar-con}\\
     H_V^a:&=& \D_b(K^{ab} -K \;^{(3)}q^{ab}) \nn \\
              &=&  \D_b P^{ab}=8\pi T^{bc} T_c \;^{(3)} q^a\;_b \label{Vector-con}
\eea
where $P^{ab}:= K^{ab} - K \;^{(3)}q^{ab}$.

 If the dominate energy condition is satisfied, it turns out that
$H$ has to be a space-like hypersurface \cite{Ashtekar02}. The
unit time-like vector that is normal to $H$ is denoted by $T^a$
and the unit space-like vector that orthogonal to the two sphere
and tangent to $H$ is denoted by $R^a$. In order to study them in
terms of Newman-Penrose quantities, they can be defined by using
the null normals $\l^a$ and $n^a$. Therefore,
\bea T^a &=& \frac{1}{\sqrt{2}}(\l^a + n^a),\;\; R^a
=\frac{1}{\sqrt{2}}(\l^a - n^a)\eea
where $T^aT_a = 1$, $R^a R_a=-1$. 
The four metric has the form
\bea g^{ab} &=& n^a \l^b + \l^a n^b + \;^{(2)}q^{ab}\\
 &=& T^a T^b - R^a R^b + \;^{(2)}q^{ab}. \eea
The three metric $^{(3)} q^{ab}$ that is intrinsic to dynamical
horizon $H$ is
\bea ^{(3)} q^{ab} = g^{ab} - T^a T^b
                   =  \,^{(2)} q^{ab} - R^a R^b. \eea
The two metric $^{(2)} q^{ab}$ that is intrinsic to the cross
section two sphere $S$ is
\bea ^{(2)} q^{ab} = \,^{(3)} q^{ab} + R^a R^b
   = -(m^a \bar m^b + \bar m^a m^b).\eea
The induced covariant derivative on H can be defined in terms of 4-dimensional covariant derivative $\nabla_a$ by
\bea \D_b V_a:= \,^{(3)}q_b\,^{c} \;^{(3)}q_a\,^{d} \nabla_c V_d,
\eea
so the three dimensional Ricci identity is then given by
\bea  ^{(3)} R_{abc}\;^d w_d = - [\D_a,\D_b] w_c.\eea
The induced covariant derivative on cross section $S$ can be defined in terms of 4-dimensional covariant derivative $\nabla_a$ by
\bea ^{(2)}\D_b V_a:= \,^{(2)}q_b\,^{c} \;^{(2)}q_a\,^{d} \nabla_c V_d.
\eea
The extrinsic three curvature $K_{ab}$ on $H$ is
\bea K_{ab} = \, ^{(3)}q_{(a}\;^c \;^{(3)}q_{b)}\;^d \nabla_c T_d
 = \nabla_a T_b - T_a \;^{(3)}a_b \nn\eea
where
  $^{(3)} a_b =T^c \nabla_c T_b.$
One can also introduce the  extrinsic two curvature $^{(2)}K_{ab}$ on $S$ by
 \bea ^{(2)}K_{ab} = \,  ^{(2)}q_{(a}\;^c \;^{(2)}q_{b)}\;^d \D_c R_d
 = \D_a R_b + R_a\;^{(2)} a_b, \nn\eea
 where
 $ ^{(2)} a_b = R^c \D_c R_b = R^{(c} \, ^{(2)}q_b\,^{d)} \nabla_c R_d.$
After a straightforward but tedious calculation, we can write the
extrinsic curvature in terms of NP spin coefficients. Here we
present the general extrinsic three curvature and two curvature
without any assumption of gauge conditions.
The extrinsic three curvature $K_{ab}$ is
\bea K_{ab} &=& \; ^{(3)} q_{(a} \,^c \nabla_c T_{b)} = \; ^{(2)}
q_{(a}\,^c \nabla_c T_{b)}
- R_{(a} R^c \nabla_c T_{b)} \nn\\
 &=& A ^{(2)}q_{ab}  + S_{ab} + 2 W_{(a} R_{b)} + B R_a R_b\eea
where
\bea
S_{ab}&=& \frac{1}{\sqrt{2}}[(\bar\sigma-\lambda)m_a m_b + C. C.], \nn\\
A &=& -\frac{\Re \rho-\Re \mu}{\sqrt{2}},\nn\\
W_a &:=& -^{(2)} q_{a} \,^c K_{cb} R^b \nn\\
&=& \frac{1}{4}[\bar\kappa+\nu-\bar\tau-\pi-2(\a+\bar\b)] m_a + C.C. , \nn\\
B &=& - \sqrt{2}(\Re \epsilon- \Re \gamma), \nn\eea
and $C. C.$ denotes the complex conjugate terms.
The extrinsic two curvature $^{(2)}K_{ab}$ is
\bea ^{(2)}K_{ab} & =&  {^{(2)}} q_{(a} \,^c \D_c
R_{b)}
= {^{(2)}} q_{(a}\,^c {^{(3)}} q_c\,^d {^{(3)}} q_{b)}\,^e \nabla_d R_e \nn\\
&=& {^{(2)}} q_a\,^d  {^{(2)}} q_b\,^e  \nabla_d R_e
= \half\; ^{(2)}K ^{(2)}q_{ab} +\, ^{(2)}S_{ab} \nn\eea
where
\bea ^{(2)}K &=& - {\sqrt{2}}(\Re \rho+\Re \mu),\label{twoK}\\
^{(2)}S_{ab}&=&\frac{1}{\sqrt{2}} (\bar\sigma+\lambda)m_a m_b + C.
C.\eea
The calculation of two acceleration $^{(2)} a_a$ yields
 \bea ^{(2)} a_a &=& R^b \D_b R_a
                = R^{(c} \, ^{(2)}q_a \,^{d)} \nabla_c R_d
                 = C m_a + \bar C \bar m_a, \nn
 \eea where
 \bea
           C = - \frac{1}{4}(\bar\kappa- \nu+\pi-\bar\tau),
 \eea
%
so the two acceleration is tangent to $S$.

We now perform 2+1 decomposition to study the various quantities
on $H$. The curvature tensor intrinsic to $S$ is
given by
\bea -^{(2)}R_{abc}\;^d &=& -^{(2)}q_a\,^f \,^{(2)}q_b\,^g
\,^{(2)}q_c\,^k \,^{(2)}q_j\,^d \,^{(3)} R_{fgk}\,^j \nn\\
&&-\; ^{(2)}K_{ac} \;^{(2)} K_b\,^d  +\; ^{(2)}K_{bc}
\;^{(2)}K_a\,^d, \nn \eea
which is the Gauss-Codacci equation. This leads to the relation
between the scalar three curvature $^{(3)} R$ and the scalar two
curvature $^{(2)} R$
\bea - ^{(3)} R = - ^{(2)} R -  ^{(2)} K^2 + ^{(2)}K_{ab}
^{(2)}K^{ab} -2 \D_a \a^a \label{scalar23}\eea
where $\a^a:= R^b \D_b R^a - R^a \D_b R^b= \;^{(2)} a^a - R^a \;
^{(2)}K  $.

From (\ref{scalar23}), we obtain the Einstein tensor on $H$
 \bea - 2\; ^{(3)}G_{ab} R^a R^b = - ^{(2)}R +^{(2)}K^2 - ^{(2)}K_{ab}\;
^{(2)}K^{ab}.\eea
The expansion of the out going tetrad $\l^a$ can be calculated to yield
 \bea \Theta_{(\l)}:=-\half(\rho+\bar\rho) =\frac{1}{2 \sqrt{2}}[K + \;^{(2)}K + B].\eea
Now we use the following relations (\ref{19})-(\ref{22}) to calculate $H_S + 2 R_a
H^a_V$, where $H_S$ and $H^a_V$ are the scalar and vector
constraints defined in (\ref{Scalar-con}) and (\ref{Vector-con}).
 \bea      &&K = 2 A- B,\label{19}\\
    && ^{(2)}K =-K-B+2\sqrt{2}\Theta_{(\l)}
             =-2A+2\sqrt{2}\Theta_{(\l)},\\
 &&K_{ab}K^{ab} = 2 A^2 + S_{ab}S^{ab} -2 W_a W^a +  B^2,\\
 &&^{(2)}K_{ab} \;^{(2)}K^{ab} = \half \;^{(2)}K^2 +\;
 ^{(2)}S_{ab}\;^{(2)}S^{ab}\label{22}.
     \eea
From the momentum constraint (\ref{Vector-con}) and use integration
by parts, we get
\bea  R_b \D_a P^{ab} = \D_a \b^a - P^{ab} \D_a R_b\eea
where $\b^a:= K^{ab} R_b - K R^a$. Thus,
\bea \c^a :=\a^a+\b^a = R^b \D_b R^a - W^a - 2 \sqrt{2}
\Theta_{(l)} R^a.\eea
For a general space-time, the matter energy flux can be calculated
as following
\bea H_s &+& 2 R_a H^a_V = \; ^{(3)}R + K^2 -K_{ab}K^{ab}+ 2 R_a \D_b P^{ab} \nn\\
&=&^{(2)}R + ^{(2)}K^2 -^{(2)}K_{ab}\;^{(2)}K^{ab} + K^2
-K_{ab}K^{ab} \nn\\
&&- 2 P^{ab} \D_a R_b
+ 2 \D_a \c^a     \;\;\; \textrm{(Use (\ref{scalar23}))}\nn\\
&=& ^{(2)}R  - \sigma_{ab}\bar\sigma^{ab} + 2 W_a W^a -2W^a \,
^{(2)}a_a \nn\\
&&+ 4 \Theta_{(\l)} (\Theta_{(\l)}- \sqrt{2} B) +2 \D_a \c^a.
\nn\eea
By applying the 2+1 decomposition on the covariant derivative $\D$ and using integration by parts, we have
\bea &&2 \D_a \c^a= 2  \D_a (\, ^{(2)} a^a - W^a - 2\sqrt{2} \Theta_{(\l)} R^a) \nn\\
                &=& 2 ( \, ^{(2)} a^a \, ^{(2)} a_a - W^a \, ^{(2)} a_a - \D_a \Theta_{(\l)} R^a + \half \, ^{(2)}
                K\Theta_{(\l)}),
                \nn\eea
where the term $^{(2)}\D_a (^{(2)} a^a - W^a)$ has been discarded since it will vanish due to the integration over a compact two surface $S$, and then
\bea &&2(W_a W^a - W_a \, ^{(2)} a^a + \D_a \c^a) \nn \\
&=& 2 (W_a -\, ^{(2)} a_a)(W^a  -\, ^{(2)} a^a) - 2 \D_a
\Theta_{(\l)} R^a - \, ^{(2)} K \Theta_{(\l)}.  \nn\eea
Here we can define
 \bea\zeta_a &:=& W_a  -\, ^{(2)} a_a= -\sqrt{2}\,
^{(2)} q_a^{(d} R^{c)} \nabla_c \l_d \nn\\
&=& \half [\bar\kappa-\bar\tau -(\a+\bar\b)] m_a + C. C.\eea
Finally, we get
 \bea H_s &+& 2 R_a H^a_V = \, ^{(2)}R -
\sigma_{ab} \sigma^{ab} +
2 \zeta_a \zeta^a \nn\\
&-& 2 \D_a \Theta_{(\l)} R^a + \Theta_{(\l)} (- \, ^{(2)} K
+4\Theta_{(\l)} - 4 \sqrt{2} B ) \nn\eea
where
 \bea \sigma_{ab} &=& \frac{1}{\sqrt{2}} (S_{ab} +\,^{(2)}
S_{ab}) =  \sqrt{2} \bar\sigma m_a m_b + C.C. \nn\eea
is the shear of null normal $\l^a$. This equation is completely
general. On the dynamical horizon, the outgoing expansion $\Theta_{(\l)}$
vanishes. It then becomes\footnote{Our expression has some minus
sign different from Ashtekar's expression because of convention.}
\bea F_{\textrm{matter}}{=}\frac{1}{16\pi} \int_{\Delta H} N (\,^{(2)}
R -\sigma_{ab} \sigma^{ab} + 2\zeta_a \zeta^a) d^3 V.
\label{startmatter}\eea
If the gauge condition
\bea\kappa \hat{=} \pi-\bar\tau \hat{=} \pi-(\a
+ \bar\b)\hat{=}0\eea
is satisfying, where $\hat{=}$ denotes the equating on DH, then $\zeta_a = -\pi m_a + C. C.$ and $\zeta_a
\zeta^a = -2 \pi\bar\pi$. So the flux formula in terms of NP in this gauge is
\bea F_{\textrm{matter}} {=}\frac{1}{16\pi} \int_{\Delta H} N (\,^{(2)}
R -4 \sigma\bar\sigma -4\pi\bar\pi) \, d^3 V. \label{AKflux}\eea

\subsection{Angular momentum flux and energy fluxes}

By contracting the vector constraint $H^a_V$ with the rotational vector
field $\psi^a$, which is tangential to $S$, we can obtain angular momentum of a black hole.
Then we integrate the resulting equation over the region of
$\Delta H$ and use the integration by parts together with the
identity $\L_\psi \;^{(3)}q_{ab} = 2 \D_{(a} \psi_{b)}$. It leads
to
\bea -d J &=& J_{S_1} - J_{S_2}\nn\\
 &=&\frac{1}{8\pi} \oint_{S_2} K_{ab}
\psi^a R^b d S - \frac{1}{8\pi} \oint_{S_1} K_{ab} \psi^a R^b
d S \nn\\
&=& \int_{\Delta H}  (T_{ab} T^a \psi^b + \frac{1}{16\pi} P^{ab}
\L_\psi \;^{(3)}q_{ab}) d^3 V. \label{J1J2}\eea
The angular momentum associated with cross-section $S$ is
$ J_S^\psi = -\frac{1}{8\pi} \oint_S K_{ab} \psi^a R^b d S $
where $\psi^a$ need not be an axial Killing field. The
\textit{flux of angular momentum} due to matter fields
$F^\psi_{\textrm{matter}}$ and gravitational waves $F^\psi_{\textrm{grav}}$ are
\bea F^\psi_{\textrm{matter}} &=& - \int_{\Delta H} T_{ab} T^a \psi^b d^3 V,\\
F^\psi_{\textrm{grav}} &=& - \frac{1}{16\pi}\int_{\Delta H} P^{ab} \L_\psi
\;^{(3)}q_{ab}  d^3 V, \eea
and the balance equation
$ J_{S_2}^\psi - J_{S_1}^\psi = F^\psi_{\textrm{matter}} + F^\psi_{\textrm{grav}}$,
which describes the difference of angular momentum between two
cross section, is due to the matter radiation and gravitational
radiation.


Each time evolution vector $t^a$ defines a horizon energy
$E^t_\Delta$. From equation (\ref{startmatter}), we find the total
energy flux is the combination of the matter flux and
gravitational flux
 \bea F_{\textrm{matter}} + F_{\textrm{grav}} = \frac{1}{16\pi}\int_{\Delta H} N  \,^{(2)} R
\;d^3 V \label{ch3-flux}\eea
where the matter flux is equation (\ref{matterflux}) and the
gravitational flux is
\bea F_{\textrm{grav}} = \frac{1}{16\pi}\int_{\Delta H} N (\sigma_{ab}
\sigma^{ab} - 2\zeta_a \zeta^a) d^3 V.\eea
If we use the gauge conditions in (\ref{AKflux}), we then have
\bea F_{\textrm{grav}} = \frac{1}{4\pi}\int_{\Delta H} N (|\sigma|^2
+|\pi|^2) d^3 V. \label{ch3-grav-flux}\eea
The matter flux expression (\ref{matterflux}) of Vaidya
solution would be
\bea F_{\textrm{matter}} :&=& \int_H T_{ab} T^a \l^b N d^3 V
               = \frac{1}{4\pi} \int \Phi_{00} N d^3 V \eea
where we use $4 \pi T_{ab} \l^a \l^b = \Phi_{00}^0$. The total flux
of Ashtekar-Krishnan then becomes
\bea F_{\textrm{total}} = \frac{1}{4\pi} \int [ |\sigma|^2
+|\pi|^2+ \Phi_{00}] N d^3 V.\label{Ash-ii} \eea
Further, the integral of $N\; ^{(2)} R$ can be written as
\bea \int_{\Delta H} N \;^{(2)} R  d^3 V =\int^{R_2}_{R_1} d R
\oint \;^{(2)} R d^2 V
= 8\pi(R_2-R_1)\nn\eea
where $R_1$ and $R_2$ are the radii of the horizon at the boundary
cross-sections. For a \textit{rotating non-spherical symmetric
dynamical horizon}, we find the relation of the change in the
horizon area in the dynamical processes can be written as
\bea &&\int \frac{d R}{2} =\frac{(R_2-R_1)}{2}= \frac{1}{16\pi
}\int_{\Delta H} N \;^{(2)} R \;d^3
V \nn \\
&=& F_{\textrm{matter}} + F_{\textrm{grav}}  \label{r2r1}\\
&=& \int_{\Delta H} T_{ab} T^a \xi^b d^3 V + \frac{1}{16\pi
}\int_{\Delta H} N (\sigma_{ab}\sigma^{ab} - 2 \zeta_a\zeta^a) d^3
V. \nn\eea
Hence, from this equation we can relate the black hole area change
with energy and angular momentum change.  This gives a more
general black hole first law in a dynamical space-time. If we now
define the \textit{effective surface gravity} \cite{Ashtekar02} as
\bea \kappa_R := \frac{1}{2 R}, \label{ESG}\eea
then the area of horizon is $A = 4\pi R^2$ and the differential of
the area is $d A = 8\pi R d R$, therefore
\bea \frac{\kappa_R}{8\pi} d A = \half d R. \eea

For the time evolute vector $t^a= N \l^a -\Omega \psi^a = \xi^a-
\Omega \psi^a $,  the difference of the horizon energy $E^t_S$ can
be expressed as  \cite{Ashtekar02}
\bea  d E^t &=& E^t_{S_2} - E^t_{S_1} \nn\\
&=& \int T_{ab} T^a t^b d^3 V
+ \frac{1}{16\pi}\int N (\sigma_{ab}\sigma^{ab} - 2
\zeta_a\zeta^a) d^3 V \nn\\
&-& \frac{1}{16\pi } \int_{\Delta H} \Omega P^{ab} \L_\psi q_{ab}
d^3 V. \label{A5.7}\eea
By using (\ref{J1J2}) together with the linear combination of
\bea \int \frac{d R}{2}  &=& \int_{\Delta H} T_{ab} T^a \xi^b d^3
V
\\ &+& \frac{1}{16\pi} \int_{\Delta H} N (\sigma_{ab}\sigma^{ab} - 2
\zeta_a\zeta^a) d^3 V, \nn\eea
we can obtain a \textit{generalized black hole first law for
dynamical horizon}
\bea \frac{\kappa_R}{8\pi} d A + \Omega d J = d E^t.\eea
%

\section{Asymptotic expansion for a spinning dynamical horizon}\label{AEDH}


\subsection{Frame setting and gauge choice}

We choose the incoming null tetrad $n_a=\nabla_a v$ to be the gradient of
the null hypersurface $v=const.$ We then have $g^{ab} v_{,a} v_{,a}
=0$. It gives us the gauge conditions $\nu=\mu-\bar\mu=
\gamma+\bar\gamma=\bar\a+\b -\bar\pi=0$. Then we further choose $n^a$ flag plane parallel, it implies $\c=0$. For the setting of outgoing null tetrad $\l$, we first choose $\l$ to be a geodesic and use null rotation type III to make $\epsilon-\bar\epsilon=0$.  We choose $m,
\bar m$ tangent to the cross section $S$, and thus $\rho\hat{=}\bar\rho, \pi\hat{=}\bar
\tau$ \footnote{This also implies $\omega\hat{=}0$. Kerr solution preferred gauge on horizon is $\mu\hat{=}\bar\mu, \pi\hat{=}\a+\bar\b\hat{=}\bar\tau$. }.
From the boundary conditions of a spinning DH ( see eq (\ref{DHgc})),  recall
\bea \rho \hat{=} 0, \pi\hat{\neq}  0, \sigma\hat{\neq}0. \eea
on DH. We summarize our gauge choices and boundary conditions
\bea \kappa=\epsilon-\bar\epsilon=\nu=\mu-\bar\mu =\c = \pi-\a-\bar\b=0,\nn\\
\rho\hat{=}0, \pi\hat{=}\bar\tau. \label{CDRD} \eea
 In order to preserve orthonormal relations,
%
%
we can choose the tetrad as
\be\ba{l} \l^a=(1, U , X^2, X^3),\;
          n^a=(0,-1,0,0),\nn\\
          m^a=(0,0, \xi^2,\xi^3). \nn\ea\ee
in the Bondi coordinate $(v, r, x^2, x^3 )$.

Now we make a coordinate transformation to a new comoving coordinate $(v, r', x^2, x^3 )$  where $r'=r-R_\Delta(v)$ and $R_\Delta(v)$ is radius of a spinning DH.  Here
\be\ba{l} \l^a=(1, U - \dot R_\Delta, X^2, X^3),\;
          n^a=(0,-1,0,0),\nn\\
          m^a=(0,0, \xi^2,\xi^3), \nn\ea\ee
where $\dot R_\Delta(v)$ is the rate of changing effective radius of DH. From this
coordinate, we may see that the dynamical horizon is a spacelike or null
hypersurface. Here, tangent vector of DH is $R^a= \l^a - \dot R_\Delta n^a \hat{=} \frac{\partial}{\partial v}$ where $\dot R_\Delta\geq 0$. Therefore, it implies $R^a R_a\leq 0$ and $U, X^k =O(r')$.

\subsection{The peeling properties and falloff of the Weyl scalars}

Since we use $\kappa=\nu=0, \sigma\neq 0, \lambda\neq 0$, then we have
\bea (\bar\d-4\a+\pi)\Psi_0 -(D-2\epsilon)\Psi_1 \hat{=}0,\\
(\Delta+\mu)\Psi_0 -(\delta-4\bar\pi-2\b)\Psi_1\hat{=}3\sigma\Psi_2,\\
(D+2\epsilon)\sigma \hat{=}\Psi_0,\\
(D+4\epsilon)\Psi_4 -(\bar\delta+4\pi+2\a)\Psi_3\hat{=}-3\lambda\Psi_2,\\
(\delta+4\b-\tau)\Psi_4 -(\Delta+4\mu)\Psi_3\hat{=}0,\\
(\Delta+2\mu)\lambda\hat{=}-\Psi_4.
\eea
in vacuum.  Therefore, one can set $\Psi_1\hat{=} \Psi_3\hat{=}0$ as peeling properties for a spinning DH. This is a  similar with perturbation method and one may refer to p. 175 and p. 180 in \cite{Chandrasekhar}.

The falloff of the Weyl scalars is
algebraically general (this is a more general setting than \cite{Wu2007} and \cite{WuWang-PRD-2009}) on DH where
\be\ba{l} \Psi_1=\Psi_3=O(r'), \Psi_0=\Psi_2=\Psi_4 =O(1).
\ea\ee
By considering Vaidya solution as our compared basis for matter field part, the falloff of the Ricci spinor components are
\bea \Phi_{00}=O(1),
\Phi_{22}=\Phi_{11}=\Phi_{02}=\Phi_{01}=\Phi_{21}=O(r').\eea

\subsection{From the radial equations}

 \bea
\mu &=& \mu_0+ ({\mu_0}^2+\lambda_0\bar\lambda_0) r' + O(r'^2), \nn \\
\lambda &=&  \lambda_0 + (2\mu_0\lambda_0+  \Psi^0_4 ) r' + O(r'^2), \nn\\
\a     &=& \a_0 +[\lambda_0(\bar\pi_0 +\b_0)  +\a_0\mu_0 ] r'+ O(r'^2), \nn\\
\b     &=& \b_0 + [\mu_0\bar\pi_0 +\b_0 \mu_0+ \a_0\bar\lambda_0 ] r'+ O(r'^2), \nn\\
\rho   &=&  [\Psi^0_2- \bar\eth_0\bar\pi_0 + \pi_0\bar\pi_0 +
\sigma_0\lambda_0]r' + O(r'^2), \nn\\
\sigma &=& \sigma_0 + (-\eth_0\bar\pi_0-\bar\pi_0^2+\mu_0 \sigma_0)r' + O(r'^2), \nn\\
\pi   &=&  \pi_0 + [2\mu_0\pi_0 + 2 \lambda_0  \bar\pi_0 ] r'+ O(r'^2), \nn\\
\epsilon &=& \epsilon_0 + [ 2\a_0 \bar\pi_0 +2 \b_0\pi_0 + \bar\pi_0\pi_0 ]   r' + O(r'^2), \nn\\
\xi^k &=& \xi^{k0} +[\bar\lambda_0\bar\xi^{k0}+ \mu_0 \xi^{k0}] r' + O(r'^2), \nn\\
U &=& 2 \epsilon_0 r' +  O(r'^2), \nn\\
X^k &=&  2(\pi_0\xi^{k0} + \bar\pi_0\bar\xi^{k0})r'+ O(r'^2).\nn\eea
\bea
     \Psi_0 &=& \Psi_0^0 + (\mu_0\Psi_0^0  - 3 \sigma_0 \Psi_2^0+ \bar\lambda_0\Phi_{00}^0) r' + O(r'^2),\nn\\
     \Psi_1 &=& ( - \delta_0 \Psi_2^0+3\bar\pi_0\Psi_2^0) r' +
      O(r'^2), \nn\\
     \Psi_2 &=& \Psi_2^0 + (3\mu_0 \Psi_2^0  - \sigma_0\Psi^0_4) r'+O(r'^2),\nn\\
     \Psi_3 &=&  [ -\delta_0\Psi_4^0 + (\bar\pi_0- 4 \b_0)\Psi_4^0] r' + O(r'^2).\nn
\eea
\bea
\Phi_{00} &=& \Phi_{00}^0 + 2 \mu_0 \Phi_{00}^0 r' + O(r'^2),\nn\\
\Phi_{11}&=& \half \dot R_\Delta \Phi_{22}^0 r'^2 + O(r'^3), \nn\\
\Phi_{01}&=& -\dot R_\Delta \Phi_{12}^0 r'^2 + O(r'^3),\nn\\
\Phi_{22}&=& \Phi_{22}^0 r'^2+ O(r'^3),\nn\\
\Phi_{12}&=& \Phi_{12}^0 r'^2+ O(r'^3),\nn\\
\Phi_{02}&=& \Phi_{02}^0 r'^2+ O(r'^2).\nn
 \eea

\subsection{From the non-radial equations }

The following equations refer to the equation numbers from p.
45-p. 47 in \cite{Chandrasekhar}. We re-label (304)-(306) as (NC1), (NC2), (NC3).
\be\ba{lll}

\textbf{(NC1)} &\delta_0 \epsilon_0=0,\\
 \textbf{(NC2)} &  \dot P = \dot R_\Delta [\mu_0 P+\bar\lambda_0 \bar P] +\sigma_0 \bar P, \\
 \textbf{(NC3)} & \bar P\up{\bar c}\nabla \ln P = \a_0 -\bar\b_0, \ea\ee
where
\bea P(v,x^k):=\xi^{20}=-i\xi^{30}, P\up{c}\nabla:=\delta_0, \up{c}\nabla:=\frac{\partial}{\partial x^2} +i\frac{\partial}{\partial x^3}. \nn\eea We relabel (a), (b), (c), (g), (d), (e), (h), (k), (m), (l) as (NR1), (NR2), (NR3), (NR4), (NR5), (NR6), (NR7), (NR8), (NR9), (NR10).
\be\ba{lll}
%

 \textbf{(NR1)} &-\dot R_\Delta (\Psi^0_2 - \bar\eth_0\bar\pi_0 +
\pi_0\bar\pi_0 + \sigma_0\lambda_0)
= \sigma_0 \bar\sigma_0 + \Phi_{00}^0,\\
\textbf{(NR2)} & \dot \sigma_0= \dot R_\Delta [-\eth_0
\bar\pi_0 -\bar\pi_0^2+ \sigma_0 \mu_0]+ 2 \epsilon_0 \sigma_0 +\Psi_0^0, \\
\textbf{(NR3)} & \dot\pi_0 = \dot R_\Delta [2\mu_0\pi_0+2\lambda_0\bar\pi_0] + 2 \bar\sigma_0\bar\pi_0,\\
\textbf{(NR4)} & \dot\lambda_0 =\dot
R_\Delta[2\mu_0\lambda_0 +\Psi_4^0] +\bar\eth_0\pi_0+\pi_0^2
\\& - 2 \lambda_0 \epsilon_0+ \mu_0\bar\sigma_0,\\
\textbf{(NR5)} & \dot \a_0 =\dot R_\Delta[\a_0\mu_0+\lambda_0(\bar\pi_0
+\b_0)] +\bar\sigma_0\b_0,\\
 \textbf{(NR6)} & \dot \b_0 =\dot R_\Delta [\a_0 \bar\lambda_0 +
\mu_0(\bar\pi_0+\b_0) ]+\sigma_0(\a_0 + \pi_0),\\
\textbf{(NR7)}  & \dot \mu_0  = \dot R_\Delta (\mu_0^2 +
\lambda_0\bar\lambda_0) +\eth_0\pi_0 +\pi_0\bar\pi_0 \\
&+ \sigma_0\lambda_0 -2 \mu_0 \epsilon_0 + \Psi_2^0, \\
  \textbf{(NR8)} &  \bar\eth_0\sigma_0 =0, \\
 \textbf{(NR9)} & \delta_0 \lambda_0 -\bar\delta_0  \mu_0 = \mu_0
\pi_0 + \lambda_0 (\bar\a_0 -3\b_0) ,\\
  \textbf{(NR10)} & \eth_0 \pi_0 - \bar\eth_0 \bar\pi_0 = 2 \Im \eth_0 \pi_0=-2
\Im (\lambda_0\sigma_0) -2 \Im \Psi^0_2,\\
\textbf{(NR10)} & \Re\Psi_2^0 = -\Re [\delta_0(\a_0-\bar\b_0)] -
\Re (\lambda_0\sigma_0)\\
& + (\bar\a_0 -\b_0)(\a_0-\bar\b_0). \nn
 \ea\ee
The following equations refer to the equation numbers (a), (b), (c), (d) on p. 49
of \cite{Chandrasekhar}. Here, we relabel (a), (b), (c), (d) as (NB1), (NB2), (NB3), (NB4).
\be\ba{llll}     %
      \textbf{(NB1)} &  -\bar\delta_0\Psi_0^0 + (4\a_0-\pi_0)\Psi_0^0 -\dot R_\Delta[-\delta_0\Psi_2^0+3\bar\pi_0\Psi_2^0]  \\
    &= -\delta_0 \Phi_{00}^0 + \bar\pi_0\Phi_{00}^0, \nn\\
      \textbf{(NB2)} &  \dot \Psi_2^0 =\dot R_\Delta (3\mu_0 \Psi_2^0 - \sigma_0\Psi^0_4 )-\lambda_0\Psi^0_0+\mu_0 \Phi_{00}^0, \nn\\
        \textbf{(NB3)} & -\bar\delta_0 \Psi_2^0 -3\pi_0 \Psi_2^0=\dot R_\Delta(-\delta_0 \Psi_4^0+(\bar\pi_0-4\b_0)\Psi_4^0) , \nn\\
       \textbf{(NB4)} & \dot \Psi_4^0 = \dot R_\Delta \Psi_4^1
     -3\lambda_0\Psi^0_2  - 4 \epsilon_0 \Psi_4^0. \nn
 \ea\ee

\subsection{Compatible constant spinor conditions for a rotating dynamical horizon} \label{rotateconstant}
In this section, we adopt a similar idea of Bramson's asymptotic
frame alignment for null infinity \cite{Bramson75a} and apply it
to set up spinor frames on the quasi-local horizons. We define the
spinor frames as
\bea Z_A\,^{\underline{A}} = (\lambda_A, \mu_A) \eea
where
$ \lambda_A = \lambda_1 o_A -\lambda_0 \iota_A$, $\mu_A = \mu_1
o_A - \mu_0 \iota_A $.
We expand $\lambda_1, \lambda_0$ as
\bea \lambda_1= \lambda_1^0(v,\theta,\phi) +
\lambda_1^1(v,\theta,\phi) r' + O(r'^2),\\
 \lambda_0=
\lambda_0^0(v,\theta,\phi) + \lambda_0^1(v,\theta,\phi) r' +
O(r'^2). \eea
Here $\lambda_1$ is type $(-1,0)$ and $\lambda_0$ is type $(1,0)$.

Firstly,we require the frame to be parallely transported
along the outgoing null normal $\l^a$.
\be\ba{l} \lim_{r'\to 0} D Z_A\,^{\underline{A}}=0.\\
\Rightarrow \l^a \nabla_a (\lambda_1 o_A -\lambda_0 \iota_A) =0.
 \ea\ee
Then it gives the condition $\thorn_0 \lambda^0_0=0$ on DH. The compatible conditions
are:
\bea \thorn_0 \lambda^0_0 &=& 0, \Rightarrow \dot \lambda^0_0 -\epsilon_0\lambda^0_0 =0 \label{TCDdh}\\
\eth_0 \lambda^0_0 + \sigma_0\lambda_1^0&=&0,\label{DM1dh}\\
\eth_0 \lambda^0_1 - \mu_0 \lambda^0_0&=& 0,\label{DM2dh}\\
\thorn_0 \lambda^0_1&=& - \bar\eth_0\lambda^0_0.\label{extradh}
 \eea

\section{Slow rotating black hole and settings on a two sphere} \label{Slow}
\subsection{Slow rotating Kerr horizon in Bondi coordinate}\label{S-Kerr}

Kerr metric in the Eddington-Finkelstein coordinate $(v, r,\theta, \chi)$ is
\bea d s^2 &=& \frac{\Delta -a^2\sin^2\theta}{\Sigma} d v^2 -2 d v
d r \nn\\
& +& \frac{2 a \sin^2\theta (r^2 + a^2 - \Delta)}{\Sigma} d v d
\chi+ 2 a \sin^2\theta d \chi d r\nn\\
&-& \Sigma d \theta^2 -
\frac{(r^2+a^2)^2 - \Delta a^2 \sin^2\theta}{\Sigma} \sin^2\theta
d \chi^2.\eea
By changing the coordinate from
$(v, \theta, \chi)$ to $(v, \theta, \chi')$
\bea d \chi = \Omega_\Delta d v + d \chi' \label{chi-pchi}\eea
where  $\Omega_\Delta = \frac{a}{r^2_\Delta + a^2}$ is the angular
velocity on horizon and $r_\Delta$ is horizon radius of Kerr solution, we can make the term $g_{v\chi}
dv d \chi$  vanished in the 3-D metric.  The 3-D metric in the new coordinate $(v, \theta,
\chi')$ will be
\bea d s^2 &\hat{=}& - \frac{a^2 \sin^2\theta}{\Sigma_\Delta} d
v^2 + 2 \frac{a^2 \sin^2\theta} {\Sigma_\Delta} \Omega_\Delta^{-1}
d v d \chi \nn\\
&& - \Sigma_\Delta d \theta^2 -
\Omega_\Delta^{-2}\frac{a^2\sin^2\theta}{\Sigma_\Delta} d \chi^2
\\
&\hat{=}& 0\cdot d v^2 - \Sigma_\Delta d
\theta^2-\Omega_\Delta^{-2}\frac{a^2\sin^2\theta}{\Sigma_\Delta}
{d \chi'}^2\\
&\cong & 0\cdot d v^2 -r_\Delta^2 (d \theta^2+\sin\theta^2 d {\chi'}^2). \eea
Here the surface area of slow rotating Kerr is $A_{Kerr}\cong 16\pi r_\Delta^2$.

Now, we consider the case of slow rotation so that $a$ is small
and we ignore the $a^2$ terms. Thus the tetrad components in the
Bondi coordinate $(\tilde{v},
\tilde{r'},\tilde{\theta},\tilde{\phi})$ are:
\bea \l^a &=& (1, U r', 0, \frac{a}{r_\Delta\, ^2} + D r') \\
     n^a &=& (0, -1, 0, 0) \\
     m^a &=& \frac{1}{\sqrt{2} \eta_\Delta}(0,0, 1- \frac{r'}{\eta_\Delta}, \frac{- i}{\sin\theta}(1- \frac{r'}{\eta_\Delta})) \eea
where $U:=\frac{r_\Delta - M}{r_\Delta^2}$ and $D := \frac{a(2
r_\Delta - M)}{r_\Delta^4}$. The NP coefficients and Weyl tensors
are:
\bea \kappa&=&\sigma=\lambda=\nu=0,\\
 \rho&=&\frac{U(-r_\Delta+ r') r'}{(\eta_\Delta -r')(\bar\eta_\Delta -r')} \hat{=} 0,\\
 \mu &=& \frac{-r_\Delta + r'}{(\eta_\Delta -r')(\bar\eta_\Delta -r')} \hat{=} -\frac{r_\Delta}{\Sigma_\Delta}\cong -\frac{1}{r_\Delta},\\
 \pi&=&\bar\tau = \frac{i \sqrt{2} D \eta_\Delta^2 \sin\theta}{4 (\eta_\Delta -r')} \hat{=} \frac{i \sqrt{2}
 D \eta_\Delta \sin\theta }{4}\nn\\
 &\cong& \frac{i 3\sqrt{2} a}{2 r_\Delta^2},\\
 \epsilon&=& \frac{U[(r'-r_\Delta)^2 + a^2 \cos^2\theta + i a \cos\theta r']}{2[(r'-r_\Delta)^2 + a^2 \cos^2\theta]}\nn\\
  &\hat{=}& \frac{U}{2} \cong \frac{1}{4r_\Delta},\\
 \c &\hat{=}& -\frac{ i a \cos\theta}{2\Sigma_\Delta}, \c+\bar\c\hat{=}0,\\
\pi&=&\a + \bar\b,\\
\Psi_0&=&0, \Psi_1= O(r'), \\
\Im \Psi_2 &\hat{=}& - \frac{i D \cos\theta}{\Sigma_\Delta}(r_\Delta^2 + a^2\cos^2\theta -a^2\sin^2\theta), \\
\Psi_3 &\hat{=} & \frac{i \sqrt{2} \sin\theta r_\Delta \eta_\Delta}{4 \Sigma_\Delta^3}
[D \Sigma_\Delta^2 + 2 i a^2\cos^2\theta],\\
\Psi_4&=&0.
 \eea
\textbf{Remark.} In this approximate Kerr tetrad in Bondi
coordinate, the  NP coefficients satisfy
\bea \nu=\mu-\bar\mu\hat{=} \pi-\a-\bar\b\hat{=} \c+\bar\c \hat{=}\epsilon-\bar\epsilon\hat{=}0,\nn\\
\pi=\bar\tau, \rho\hat{=}\bar\rho, \mu<0. \eea
By examining the approximate Kerr tetrad in Bondi coordinates, we
found it is compatible with our frame setting for the asymptotic
expansions.

\subsection{Setting on a two sphere: on horizon cross section} \label{sphere}


To solve the coupling equations from non-radial NP equations would be rather complicated and maybe too general to yield some interesting physical results. By considering the small-tide and slow rotate of DH and consider slow rotate Kerr solution as a basis from pervious subsection, we use  two sphere conditions of DH cross section for our later calculation.  On a sphere with horizon radius $R_\Delta(v)$, one can set
\bea \mu_0= -\frac{1}{R_\Delta}. \eea
Let $P, \mu_0$ on a sphere with radius $R_\Delta$, then $P \propto \frac{1}{R_\Delta}$. From (NC2), $\dot\xi^{k0} = \dot {R}_\Delta (\mu_0 \xi^{k0} + \bar\lambda_0\bar\xi^{k0}) + \sigma_0 \bar\xi^{k0}$ which depends on the next order nonlinear effect off horizon, we obtain
\bea \lambda_0 = -\frac{\bar\sigma_0}{\dot{R}_\Delta}, \eea
and
\bea \dot P =\dot R_\Delta \mu_0 P = -\frac{\dot R_\Delta P}{R_\Delta}. \eea

Moreover, the effective surface gravity is $\tilde{\kappa}=2\epsilon_0 = \frac{1}{2R_\Delta}$, and then $\mu_0=-4\epsilon_0$ (Recall eq. (\ref{ESG})).

Check the commutation relation $[\delta_0,D_0]\lambda_0$ and $[\delta_0,D_0]\sigma_0$, it implies
\bea\ddot R_\Delta=0.\eea
This means that the horizon radius will not accelerate (no inflation). The dynamical horizon will increase with a constant speed. We note here that if the two sphere condition does not hold, then this result is no longer true.

After applying these conditions, we list the main equations that will be used in the later section
\be\ba{lll}
%
 \textbf{(NR1')} &\dot R_\Delta [-\half(\Psi^0_2 +\bar\Psi^0_2)\\ &\;\;\;\;\;+\half(\eth_0\pi_0+ \bar\eth_0\bar\pi_0) -
\pi_0\bar\pi_0 ] =  \Phi_{00}^0,\\
\textbf{(NR2')} & \dot \sigma_0= \dot R_\Delta [-\eth_0
\bar\pi_0 -\bar\pi_0^2+ \sigma_0 \mu_0]\\&\;\;\;\;\;+ 2 \epsilon_0 \sigma_0 +\Psi_0^0, \\
\textbf{(NR3')} & \dot\pi_0 = 2\dot R_\Delta \mu_0\pi_0,\\
\textbf{(NR4')} & \frac{\bar\sigma_0 \ddot{R}_0 -\dot R_\Delta\dot{\bar\sigma_0}}{(\dot R_\Delta)^2} =\dot R_\Delta \Psi_4^0 \\ &\;\;\;\;\; +\bar\eth_0\pi_0+\pi_0^2
 + 2 \frac{\bar\sigma_0 \epsilon_0}{\dot R_\Delta}- \mu_0\bar\sigma_0,\\
\textbf{(NR5')} & \dot \a_0 =\dot R_\Delta\a_0\mu_0-\bar\sigma_0\bar\pi_0,\\
 \textbf{(NR6')} & \dot \b_0 =\dot R_\Delta \mu_0(\bar\pi_0+\b_0) +\sigma_0 \pi_0,\\
\textbf{(NR7')}  & \Re\Psi_2^0 = 2 \mu_0 \epsilon_0 -\pi_0\bar\pi_0-\Re\eth_0\pi_0, \\
& \Im \Psi_2^0 =-\Im\eth_0\pi_0,\\
  \textbf{(NR8')} &  \bar\eth_0\sigma_0 =0, \\
 \textbf{(NR9')} & -2\bar\sigma_0\bar\pi_0= \dot R_\Delta \mu_0\pi_0 ,\\
& \sigma_0\bar\eth_0\pi_0=-\half\dot R_\Delta \mu_0\bar\eth_0\bar\pi_0,\\
\textbf{(NB2')} & \dot \Psi_2^0=\dot R_\Delta[3\mu_0-\sigma_0\Psi_4^0]+\bar\sigma_0\Psi_0^0+\mu_0\Phi_{00}^0,\\
\textbf{(NR1')+(NR7')} & 2\Re\eth_0\pi_0=\frac{\Phi_{00}^0}{\dot R_\Delta}+2\mu_0\epsilon_0 . \nn
 \ea\ee
%
\section{Angular momentum and angular momentum flux of a slow rotating DH} \label{J-DH}

Here we use an asymptotically rotating Killing vector $\phi^a$ for a spinning DH.  It coincides with a rotating vector $\phi^\a\hat{=}\psi^a$ on a DH and is divergent free. It implies $\Delta_a \phi^a := S_{a0}^a S_a^{b0}\nabla_{b0} \phi^{a0}=0$. Therefore,
\bea \bar m_a \delta \phi^a = -m_a\bar\delta\phi^a. \eea
Let $\phi^a=A m^a+ B \bar m^a$, we get $A=-B$. Therefore, it exists a function $f$ such that
\bea \phi^a = \bar\delta f m^a-\delta f \bar m^a,\eea
which is type $(0,0)$. Since $f$ is type $(0,0)$, therefore $\delta f=\eth f$.

By using Komar integral, the quasi-local angular momentum on a slow rotating DH is
\bea  J(R_\Delta) &=& \frac{1}{8\pi} [\oint_S  \nabla^a \phi^b d S_{ab}]|_\Delta\nn\\
                  &=&  \frac{1}{8\pi} \oint_S  \Im(\bar\pi_0\bar\eth_0 f ) d S_\Delta \;\;\textrm{(use integration by part)}\nn\\
                  &=&-\frac{1}{4\pi} \oint_S  f \Im \eth_0 \pi_0 d S_\Delta\nn\\
                  &=& -\frac{1}{4\pi} \oint_S  f \Im \Psi_2^0 d S_\Delta.\eea
From (NB2'), we get $\Im \dot\Psi_2^0 = 3 \frac{\dot{R}_\Delta}{R_\Delta} \Im \eth_0 \pi_0 =-3 \frac{\dot{R}_\Delta}{R_\Delta} \Im \Psi_2^0$. Together with
$\frac{\partial}{\partial v} d S_\Delta= 2\frac{\dot{R}_\Delta}{R_\Delta} d S_\Delta$, the angular momentum flux for a slow rotating DH is
\bea \dot J(R_\Delta) = -\frac{1}{4\pi} \oint_S (\dot f -\frac{\dot R_\Delta}{R_\Delta} f) \Im \Psi_2^0 d S_\Delta\nn\\
= \frac{1}{4\pi} \oint_S \Im[ (\dot f -\frac{\dot R_\Delta}{R_\Delta} f) \eth_0\pi_0] d S_\Delta. \eea
We note that from $\frac{d}{d v}$ (NR7'), it yields the same result. Here if $\pi_0\neq 0$ and $f(v,\theta,\phi)=G(\theta,\phi)R_\Delta(v)$, then $\dot J(R_\Delta)=0$. It then returns to the stationary case. If $\pi_0 =0$, i.e., $\Im\Psi_2^0=0$, then $J$ and $\dot J=0$. It then returns to the non-rotating black hole.

For IH, by using integration by parts and Penrose volume I, p. 281, we have
\bea \oint \omega_a \phi^a = \oint f \Im \eth \pi = \oint f \Im \Psi_2.\eea
For DH, we have
\bea K_{ab} \phi^a R^b = 2 W_{(a}R_{b)} \phi^a R^b = -W_a \phi^a\\
= -\pi m_a \phi^a + C. C. = -\pi \eth f +\bar\pi\bar\eth f.\eea

\section{The quasi-local energy-momentum and flux of a slow rotating
DH}\label{EM-spinningDH}

\subsection{Mass and mass flux from Komar integral} \label{K-mass}

The asymptotic time Killing vector on a DH can be expressed as $t^a_0=\frac{\partial}{\partial v} = [\l^a+(U-\dot R_\Delta)n^a]|_\Delta$ in corotating coordinate. The Komar mass on a DH
is then
\bea M_\Delta &=& \frac{1}{8\pi}\oint_S \nabla^a t^b_0 N d S_{ab}
=\frac{1}{4\pi}\oint \epsilon_0 N d S_\Delta \nn\\
&=& \frac{1}{4\pi}\oint \frac{1}{4 R_\Delta} N d S_\Delta, \eea
where $\epsilon_0=-\mu_0/4$. This yield the same with eq. (i) from two spinor calculation when one chose $N=\lambda^0_0 \bar\lambda^0_{0'}$.

We then obtain the mass flux on a DH from Komar integral is
\bea \dot M_\Delta = \frac{1}{4\pi}\oint \frac{\dot R_\Delta}{R_\Delta^2} N d S_\Delta, \eea
and later we shall see that it agrees with eq. (\ref{NewsDH}) from two spinor method.

\subsection{Mass and mass flux from two spinor method} \label{Mass-spinor}

By using the compatible constant spinor conditions for a spinning dynamical
horizon (\ref{DM1dh}), (\ref{DM2dh}) and the results of the asymptotic
expansion, we get the quasi-local energy-momentum integral on a slow rotating
 dynamical horizon
%
\be\ba{llll} I(R_\Delta)&=&
-\frac{1}{4\pi}\oint \mu_0
\lambda_{0}^0
\bar\lambda_{0'}^0 d S_\Delta &\textrm{(i)}\\
&=&- \frac{1}{4\pi} \oint
\frac{\Re}{2\epsilon_0}[\Psi^0_2 +\delta_0 \pi_0+2\b_0\pi_0] \lambda_0^0 \bar\lambda^0_{0'} d S_\Delta. &\textrm{(ii)}\nn
\ea\ee

In order to calculate flux we need the time related condition (\ref{TCDdh}) of constant spinor of
dynamical horizon in Section \ref{rotateconstant} and re-scale it.
Then $\dot{\lambda}^0_0=0$.
It's tedious but straightforward to calculate the flux expression.
It largely depends on the non-radial NP equations and the second
order NP coefficients. By using Sec.\ref{sphere}, we substitute them back into the energy-momentum flux formula to
simplify our expression.

\textbf{From (i)} Apply time derivative to (i), and then we obtain the \textit{quasi-local
energy momentum flux for dynamical horizon}
\bea \dot I(R_\Delta) = \frac{1}{4\pi}\oint \dot \mu_0
\lambda_{0}^0 \bar\lambda_{0'}^0 d S_\Delta. \label{NewsDH} \eea
where it is always \textit{positive}. Here $\dot\mu_0$ is the
\textit{news function of DH} that always has mass gain.

 From the choice of $\mu_0 =-\frac{1}{R_\Delta}$, we
have $\dot\mu_0 =\frac{\dot R_\Delta}{R_\Delta^2}=  \frac{\dot
R_\Delta}{2} \;^{(2)} R $ where the two scalar curvature is
$^{(2)} R = \frac{2}{R_\Delta^2}$ (The metric of a two sphere with
radius $R_\Delta$ is $d \tilde{s}^2 = - R_\Delta^2 (d\theta^2 +
\sin^2\theta d\phi^2)$.). Integrate the above equation with
respect to $v$ and use $\dot \mu_0 = \dot R_\Delta {^{(2)}} R/2$,
we then have \cite{Wu2007}
\bea d I(R_\Delta) =\frac{1}{8\pi} \int \;^{(2)}R \lambda_{0'}^0
\bar\lambda_0^0 d S_\Delta d R_\Delta. \label{areaDH}\eea

\textbf{From (ii)} We first apply $\partial/\partial v$ on (NR7) to get
\bea \dot \mu_0 = \frac{\dot\Psi_2^0+\dot\delta_0\pi_0+\delta_0\dot\pi_0+2\dot\b_0\pi_0+2\b_0\dot\pi_0}{2\epsilon_0} -\frac{\mu_0\dot\epsilon_0}{\epsilon_0}
\eea

 Now, apply time derivative on (ii) and use Sec. \ref{sphere} yields
%
 \bea \dot I(R_\Delta) &=& \frac{1}{4\pi} \oint  \frac{1}{2
\epsilon_0} \{\frac{1}{ R_\Delta} \Phi_{00}^0 -2\frac{\sigma_0\bar\sigma_0}{\dot R_\Delta}( \frac{\partial}{\partial v} \ln( R_\Delta^2 \sigma_0\bar\sigma_0)) \nn\\
&&
+3\frac{\dot R_\Delta\pi_0\bar\pi_0}{ R_\Delta} \}\lambda_{0}^0
\bar\lambda_{0'}^0 d S_\Delta \label{grav-flux-DH}\eea
%
where the total energy momentum flux $F_{\textrm{total}}$ is LHS of (\ref{grav-flux-DH}) and is equal to matter flux plus gravitational flux $F_{\textrm{total}}=F_{\textrm{matter}}+F_{\textrm{grav}}$.  We can write the gravitational flux equal to the shear flux plus angular momentum flux.
\bea F_{\textrm{grav}}=F_{\sigma}+F_{J} \eea
 where the shear flux $F_{\sigma}$ is second term of RHS of (\ref{grav-flux-DH}) and the angular momentum flux $F_{J}$ is third term of RHS of (\ref{grav-flux-DH}). The coupling of the shear $\sigma_0$ and $\pi_0$ can be transform into $\pi_0$ terms by using (NR9'). Then integrate the above equation with respect to $v$, we have
%
 \bea d I(R_\Delta) &=& \frac{1}{8\pi}\int \frac{ R_\Delta}{\dot R_\Delta}\{\frac{1}{ R_\Delta} \Phi_{00}^0 -2\frac{\sigma_0\bar\sigma_0}{\dot R_\Delta}( \frac{\partial}{\partial v} \ln( R_\Delta^2 \sigma_0\bar\sigma_0)) \nn\\
&& +3\frac{\dot R_\Delta\pi_0\bar\pi_0}{ R_\Delta}\} \lambda_{0}^0 \bar\lambda_{0'}^0 d
S_\Delta d R_\Delta. \label{ch6-dI}\eea
%
where $d v = \frac{d R_\Delta}{\dot R_\Delta}$. Here we note that if one wants to observe positive shear flux $-\frac{\partial}{\partial v}\ln (R_\Delta^2\sigma_0\bar\sigma_0)\geq 0$, it implies that
\bea \dot \sigma_0 \leq 0,\eea
where $\dot R_\Delta, R_\Delta >0$ have been considered. So the shear on a spinning DH is monotonically decreasing with respect to $v$.

Recall that the total flux of Ashtekar-Krishnan (\ref{Ash-ii}), we
compare our expression with Ashtekar's expression.
%
%
If we choose $N= \lambda_{0}^0 \bar\lambda_{0'}^0$,
 then (\ref{ch6-dI}) together with (\ref{areaDH})
gives
\begin{widetext}
\bea d I(R_\Delta) &=& \frac{1}{8\pi} \int \;^{(2)}R N d S_\Delta
d R_\Delta = \frac{1}{8\pi}\int \frac{ R_\Delta}{\dot R_\Delta}\{\frac{1}{R_\Delta} \Phi_{00}^0 +2 k \frac{\sigma_0\bar\sigma_0}{\dot R_\Delta}
 +3\frac{\dot R_\Delta\pi_0\bar\pi_0}{ R_\Delta}\} N d
S_\Delta d R_\Delta  \label{DHflux} \eea \end{widetext}
where we define $\frac{\partial}{\partial v} \ln(R_\Delta^2\sigma_0\bar\sigma_0):=-k$ for the convenience. This is the relation between the change in DH area (Recall
(\ref{r2r1})) and total flux with including the matter flux and
gravitational flux.

\textbf{Shear flux:}
In the special case  $\frac{\partial}{\partial v} \ln(R_\Delta^2\sigma_0\bar\sigma_0):=-k$ where $k$ is a constant, we then have
\bea R_\Delta^2\sigma_0\bar\sigma_0 = A e^{-k v}.\eea
If $k>0$, $\sigma_0\searrow$. If $k<0$, $\sigma_0\nearrow$. Therefore, if we want to get positive gravitational flux, the shear $\sigma_0$ must decrease with time $v$ and $k>0$. On the contrary, the negative gravitational flux implies the shear must grow with time.  The negative mass loss from shear flux will make the dynamical horizon grow with time is physically  unreasonable.
Therefore, the second term of RHS in eq. (\ref{grav-flux-DH})  should be positive. This says that the shear on a spinning DH will decay to zero when time $v$ goes to infinity and the amount of shear flux $F_\sigma$ is finite.
\bea \sigma_0 \rightarrow 0, |v|\rightarrow \infty,\\
      F_\sigma< \infty.\eea
Hence a slow rotating dynamical horizon will settle down to an equilibrium state, i.e., isolated horizon at late time.

\textbf{Discussion}
\begin{enumerate}

 \item If  $\pi_0=0$ and the shear does not vanishes $\sigma_0\neq0$ we have
\bea \dot I(R_\Delta) = \frac{1}{4\pi}\oint \frac{2 R_\Delta}{\dot R_\Delta} [ \Phi_{00}^0 +2 k \sigma_0\bar\sigma_0  ] N d S_\Delta. \nn\eea
This goes back to the result of the flux of the non-rotating
dynamical horizon. When $k=\half$, it goes back to the result of non-rotating DH in \cite{WuWang-PRD-2009}.

    \item If both shear and $\pi_0$ vanishes, we have
\bea \dot I(R_\Delta) = \frac{1}{4\pi} \oint
\frac{R_\Delta\Phi_{00}^0}{\dot R_\Delta} N d S_\Delta. \nn\eea
This result can be compared with dynamical horizon of Vaidya solution.
\item Though we chose the cross section of DH to be two sphere, however, it still imply that the shear term cannot make into zero. This is because the contribution of shear  comes from the next order nonlinear effect of the equations.

\end{enumerate}

\textbf{Laws of black hole dynamics}

LHS of eq. (\ref{DHflux}) can be written as
\bea\frac{d I(R_\Delta)}{2}= \frac{\tilde{\kappa}}{8\pi} d A = \frac{d R_\Delta}{2}\eea
where $A=4\pi R_\Delta^2$. For a time evolute vector $t^a=N \l^a -\Omega \phi^a$, the difference of horizon energy $d E^t$ can be calculated as follow
\bea d E^t &=& \frac{1}{16\pi} \int \frac{ R_\Delta}{\dot R_\Delta}\{\frac{1}{R_\Delta} \Phi_{00}^0 +2 k \frac{\sigma_0\bar\sigma_0}{\dot R_\Delta}
 \} N \nn\\
 &&+[3 N \pi_0\bar\pi_0-4\frac{\Omega}{\dot R_\Delta}\Im[(\eth_0\dot f-\frac{\dot R_\Delta}{R_\Delta}\eth_0 f )\pi_0] d V\nn\eea
and the generalized black hole first law for a slow rotating dynamical
horizon is
\bea \frac{\tilde{\kappa}}{8\pi} d A +\Omega d J = d E^t.\eea

\section{Conclusions}

Since $\Psi_0,\Psi_4$ are gauge invariant quantities in a linear perturbation theory, it allows us to chose a gauge, in which $\Psi_1,\Psi_3$ vanish on DH. This choice of gauge is crucial for the coupling of the NP equations and the consequence physical interpretation. In this paper, we use a different peeling property from our earlier work \cite{WuWang-PRD-2009}\cite{Wu2007}. This leads to a physical picture that captures a collapsing slow rotating star and formation of a dynamical horizon that finally settle down to an isolated horizon at late time. Further from the peeling property, if the shear flux is positive, it excludes the possibility for a slow rotating DH to absorb the gravitational radiation from nearby gravitational sources. The mass and momentum are carried in by
the incoming gravitational wave and cross into dynamical horizon. We shall see that though it may exist outgoing wave on horizon, however, it will not change the boundary condition or make the contribution to the energy flux. A dynamical horizon forms inside the star and eat up all the incoming wave when it reaches the equilibrium state, i.e., isolated horizon.

The NP equations are simplified by using  two-sphere conditions for a slow rotating DH with small tide. 
By using the compatibility of the coupling NP equations and the asymptotic constant spinors, the energy flux that cross into a slow rotating DH should be positive. The mass gain of a slow rotating DH can be quantitatively written as matter flux, shear flux and angular momentum flux. Further, a result  comes out that the shear flux must be positive implies the shear must monotonically decay with respect to time. This is physically reasonable since black hole cannot eat infinite amount of gravitational energy when there is no other gravitational source near a slow rotating DH. We further found that the mass and mass flux based on Komar integral can yield the same result. Therefore, our results are unlikely expression dependent. For other quasi-local expressions remain the open question for the future study. It would be interesting if one can free the two-sphere conditions, then obtain the metric distorted by gravitational wave. 

\acknowledgments

YHW would like to thank Prof James M. Nester for helpful discussion. YHW would like to thank Top University Project of NCU supported by Ministry of Education, Taiwan. YHW was supported by Center for Mathematics and Theoretical Physics, National Central University, Taiwan. CHW was supported by the National Science Council of Taiwan under the grants NSC 96-2112-M-032-006-MY3 and 98-2811-M-032-014.


\end{document}